%% file: paperdraft.tex
\documentclass[12pt,a4paper]{article}

  \usepackage{color}
  \usepackage{a4wide}
  \usepackage{latexsym}
  \usepackage{epsf}
  \usepackage{amssymb,bbm}
  \usepackage{graphicx}
  \usepackage{amsmath, cite}
  \usepackage{amsmath,amssymb,amsthm}
  \usepackage{verbatim}
  \usepackage{hyperref}

\newcommand{\f}[2]{\frac{#1}{#2}}

\newcommand{\dd}{\mathrm{d}}

\newcommand{\e}{\mathrm{e}}
\newcommand{\w}{\wedge}
\newcommand{\bbm}{\left(\begin{matrix}}
\newcommand{\ebm}{\end{matrix}\right)}
\newcommand{\bea}{\begin{eqnarray}}
\newcommand{\eea}{\end{eqnarray}}
\newcommand{\be}{\begin{equation}}
\newcommand{\ee}{\end{equation}}

\newcommand{\vol}{\text{vol}}

\renewcommand{\cal}[1]{\mathcal{#1}}

\renewcommand{\d}{\textrm{d}}

\begin{document}
\numberwithin{equation}{section}

\begin{center}

\begin{flushright}
\end{flushright}
\vspace{1.5 cm}

{\LARGE {\bf Smeared antibranes polarise in AdS }}  \\

\vspace{1.5 cm} {\large  Fridrik Freyr Gautason, Brecht Truijen and Thomas Van
Riet }\\
\vspace{0.2 cm}  \vspace{.15 cm} { Instituut voor Theoretische Fysica, K.U. Leuven,\\
Celestijnenlaan 200D B-3001 Leuven, Belgium}

\vspace{0.7cm} {\small \upshape\ttfamily  ffg, brecht.truijen, thomas.vanriet @fys.kuleuven.be
 }  \\

\vspace{1.2cm}

{\bf Abstract}
\end{center}

{\small In the recent literature it has been questioned whether the local backreaction of antibranes in flux throats can induce a perturbative brane-flux decay. Most evidence for this can be gathered for D6 branes and D$p$ branes smeared over $6-p$ compact directions, in line with the absence of finite temperature solutions for these cases. The solutions in the literature have flat worldvolume geometries and non-compact transversal spaces. In this paper we consider what happens when the  worldvolume is AdS and the transversal space is compact. We show that in these circumstances brane polarisation smoothens out the flux singularity, which is an indication that brane-flux decay is prevented.  This is consistent with the fact that the cosmological constant would be less negative after brane-flux decay. Our results extend recent results on AdS$_7$ solutions from D6 branes to AdS$_{p+1}$ solutions from D$p$ branes. We show that supersymmetry of the AdS solutions depend on $p$ non-trivially.}

\setcounter{tocdepth}{2}
\newpage

\section{Introduction}

In the recent years there has been an active study of supergravity solutions that feature D$p$-branes locally surrounded by fluxes that induce a delocalised D$p$ charge density of the opposite sign to the brane charge. These different signs can be seen in the Bianchi identity:
\be
\d F_{8-p} = H\w F_{6-p} + Q\delta_{8-p}\,.\label{Bianchi1}
\ee
The solutions of interest are such that the orientation of the first form on the RHS of (\ref{Bianchi1}) is opposite to second one. For this reason we name those branes ``antibranes'' where the ``anti'' refers to the charge being opposite to the charge density in the fluxes.

Solutions with this distinctive property can be categorised into two classes:
1) Non-supersymmetric solutions with flat D$p$ worldvolume and 2) solutions with a D$p$ worldvolume that is AdS and a transversal space that is potentially compact. The AdS solutions can be supersymmetric \cite{Apruzzi:2013yva}, but do not need to be \cite{Junghans:2014wda}. 

Examples of the first kind are the non-compact geometries in which supersymmetry (SUSY) is broken by the brane, with the prime example being anti-D3 branes in the Klebanov-Strassler throat \cite{McGuirk:2009xx, Bena:2009xk, Bena:2011wh} first studied in \cite{Kachru:2002gs}. When combined with orientifolds  and quantum corrections there is a believe that it can be made into a compactification geometry for which the SUSY-breaking branes are used to uplift AdS vacua to dS vacua \cite{Kachru:2003aw} or as a base for brane inflation \cite{Kachru:2003sx}. 

Examples of the second kind are warped AdS$_7$ vacua in massive IIA SUGRA build from space-filling D6 branes and a transversal space that is conformal $S^3$. These solutions were first uncovered in \cite{Blaback:2011nz, Blaback:2011pn}, but their supersymmetry together with more details was properly understood in \cite{Apruzzi:2013yva}. These solutions provide furthermore a concrete gravity dual to six-dimensional $(1,0)$ SCFT's \cite{Gaiotto:2014lca}.

Both classes of solutions feature a peculiar property that has been the origin of an ungoing debate: due to the differences in charges, the fluxes are electromagnetically (and gravitationally) attracted to the branes in such a way that a singular flux cloud is formed around the branes \cite{DeWolfe:2004qx, Blaback:2011nz, Blaback:2012nf, Michel:2014lva}. This was first uncovered in \cite{McGuirk:2009xx, Bena:2009xk} and by now a vast literature on this exists, with a formal proof for this unavoidable singularity presented in \cite{Gautason:2013zw, Blaback:2014tfa} \footnote{Around orientifolds this phenomenon does not occur since they gravitationally repel the flux as much as it is attracted \cite{Blaback:2010sj}. This nicely fits together with orientifold compactifications were orientifolds in fluxes of opposite charge are the basic principle \cite{Giddings:2001yu}.}. Since fluxes can materialise into actual branes \cite{Kachru:2002gs} one is tempted to conclude that a singular, or large, flux pile-up leads to a quick annihilation of the flux with the antibrane, possibly making the solution perturbatively unstable \cite{Blaback:2012nf, Danielsson:2014yga} (see also \cite{Michel:2014lva} for recent comments on this). This picture is strengthened by the absence of regular solutions at finite temperature with flat worldvolume \cite{Bena:2012ek, Bena:2013hr, Blaback:2014tfa}. 

Recently this interpretation of a perturbative decay has been challenged by some good arguments \cite{Michel:2014lva, Hartnett:2015oda}. First of all it has been claimed that the singularity will get resolved by stringy corrections in such a way that the resulting flux clumping is small enough for at least a single antibrane to be meta-stable \cite{Michel:2014lva}. This can very well be correct and is currently under investigation, but it would clearly be more gratifying if the singularity can be resolved at supergravity length scales, such that the original arguments for antibrane meta-stability \cite{Kachru:2002gs} are applicable. Interestingly a resolution at sufficiently large length scales has been argued by Hartnett in \cite{Hartnett:2015oda}. The basic claim of \cite{Hartnett:2015oda} is that the nogo-theorem for finite temperature resolutions of \cite{Blaback:2014tfa} can be circumvented. Secondly a simplified trick was found to understand the local geometry of localised antibranes and it seems to indicate that a Polchinski-Strassler (PS) type of singularity resolution \cite{Polchinski:2000uf} will take place that dilutes the flux clumping strongly enough to prevent direct brane-flux decay. This is in contrast with earlier investigations of a possible PS resolution that turned out not to work \cite{Bena:2012tx, Bena:2012vz}, and according to \cite{Hartnett:2015oda} the reasons for this is the use of smeared antibranes\footnote{For the anti-D6 solutions \cite{Blaback:2011pn} no smearing was used and the absence of a PS resolution is not questioned in this particular case.} instead of localised ones. It would be interesting to verify this explicitly. In this paper we do consider smeared antibranes but we demonstrate that, when they live in AdS space, their singularities do get resolved as opposed to flat space. In that sense there is no disagreement in the literature when it comes to flat space smeared anti-Dp branes or localised anti-D6 branes: the absence of regular finite T solutions seems without doubt and perfectly in line with the `no PS-resolution' results. 
Hence perturbative flux-brane decay might very well be what is going on with the flat space solutions, but is inconsistent with the AdS solutions which, upon brane-flux decay have a less negative CC and hence more energy, so one expects those solutions to be stable.

In the case of the anti-D6 brane this has been understood by now. For the supersymmetric AdS$_7$ solutions it was found that the (anti-)D6-branes polarise into spherical D8 branes\cite{Apruzzi:2013yva,Junghans:2014wda}, which resolves the flux singularity since a charged sphere attracts the flux in a more delocalised fashion. This polarisation does not occur for the flat SUSY-breaking solutions \cite{Bena:2012tx} and we expect the latter solutions to decay perturbatively before the flux reaches the singular values. In this paper we extend this picture to the other branes and uncover that a similar story holds (up to certain subtleties): the compact AdS$_{p+1}$ solutions are such that the D$p$ branes polarise into spherical D($p+2$) branes and brane-flux decay does not occur. For D3 branes brane-flux decay can be studied very explicitly since in that case it proceeds through brane polarisation into an NS5 brane \cite{Kachru:2002gs} in a direction orthogonal to the D5 polarisation. For the other branes it is unclear what the explicit brane-flux decay mechanism is and is most likely related to a T-dual version of spherical NS5-branes.

The rest of this paper is organised as follows. In section \ref{antiD3sol} we discuss smeared anti-D3 solutions and the corresponding AdS$_4$ vacua and in section \ref{Myers} we verify whether the singularities get resolved by polarising into spherical D5 branes and whether brane-flux decay can be prevented. In section \ref{generalDpsolutions} we generalise the discussion to D$p$ branes smeared over $(6-p)$ directions and we conclude in section \ref{conclusion}. We have added various appendices of which one discusses the constraints that SUSY puts on the AdS solutions in this paper.

\section{Anti-D3 solutions}\label{antiD3sol}
In this section we describe the compact AdS$_4$ solutions build from anti-D3 branes whose RR tadpole is canceled by 3-form fluxes and non-compact anti-D3 solutions with flat worldvolumes that do not require an RR tadpole condition. The existence of the AdS$_4$ solutions found here were established in the limit of smeared antibranes in \cite{Silverstein:2004id}  (and generalised to other dimensions in \cite{Blaback:2010sj}). We have not yet made an attempt to construct the solutions with fully localised branes but we expect them to be contained in the analysis of \cite{DHoker:2007xy}.
The reader interested in checking the calculations will find use of Appendix \ref{Notation} that fixes notation and conventions.

\subsection{Ansatz}
We consider the following Ansatz for the metric:
\begin{equation}\label{metric}
 \dd s^2 = \e^{2A}\dd s^2_{\text{AdS}_4} + \e^{2B}\dd s^2_{S^3} + \e^{2C}\Bigl(\dd \rho^2 + \e^{2D}\dd \Omega^2\Bigr)\,,
\end{equation}
where $\dd s_{\text{AdS}_4}^2$, $\dd s_{S^3}^2$ and $\dd \Omega^2$ are the metrics for AdS$_4$, the three-sphere $S^3$ and two-sphere $S^2$ respectively.
These metrics are chosen such that the corresponding unwarped Ricci tensors take the canonical form
\be
\hat{R}_{\mu\nu} = \Lambda \hat{g}_{\mu\nu}~,\quad \hat{R}_{mn} = 2\hat{g}_{mn}~,\quad \hat{R}_{ij} = \hat{g}_{ij}~.
\ee
Here the hatted variables denote the unwarped quantities so that 
\be
\dd s^2_{\text{AdS}_4} = \hat{g}_{\mu\nu} \dd x^\mu\dd x^\nu~, \qquad \dd s^2_{S^3} = \hat{g}_{mn} \dd x^m\dd x^n~, \quad
\dd \Omega^2 = \hat{g}_{ij} \dd x^i\dd x^j~,
\ee
this equation also serves to fix the index conventions that we use in the following.
The warp factors $A$, $B$, $C$ and $D$ are only functions of the coordinate $\rho$ and so the Einstein equations reduce to ordinary differential 
equations for the warp factors. 
We parametrize our non-zero fluxes as follows:
\begin{align}
& F_3 = M\vol_3~,\nonumber\\
& H  = -\lambda \e^\phi\star_6 F_3~, \label{fluxes}\\
& F_5 = (1+\star_{10})\star_6\e^{-4A}\dd\alpha~.\nonumber
\end{align}
Like the warp factors, $\lambda, \alpha$ and $\phi$ are functions of the $\rho$ coordinate, whereas $M$ is a constant topological flux quantum.
The volume form $\vol_3$ is the volume form on the unit three-sphere. The equations of motion turn into a system of coupled ordinary differential equations (ODE's) and are written out explicitly in appendix \ref{secondordereq}. 
From the equation of motion for $H$ it then immediately follows that 
\be
\alpha = \lambda \e^{4A}~,
\ee
in the case where $\lambda$ is $\pm 1$ the 3-form fluxes combine into ISD or anti-ISD flux as in \cite{Giddings:2001yu} but this is not 
true in general due to the non-trivial 5-form field strength.
The Bianchi identity for $F_5$ takes the form
\be\label{F5bianchi}
\dd F_5 - H\w F_3 = N_\text{D3}\mu_3 \delta_6~,
\ee
where $\delta_6$ is a 6-form with delta function support on the worldvolume of the branes. Notice that we have added an integer $N_\text{D3}$ to allow
for a stack of D3-branes. Integrating the Bianchi identity over the internal space we obtain the tadpole cancellation condition, 
\be\label{tadpole}
\int F_3\w H = Q~,
\ee
where $Q$ denotes the total brane charge.
It is clear that even though we have included the effect of D3-branes in the Bianchi identity \eqref{F5bianchi}, the form of the metric
\eqref{metric} does not allow for fully localised branes. Indeed the branes are smeared over the three-sphere, but are localised at $\rho=0$.

\subsection{Fully smeared 3-branes}
A fully smeared limit of the solution can be obtained by replacing the delta function in equation \eqref{F5bianchi} with its integrated average
\be
\delta_6 \to \vol_6~,
\ee
where the internal volume form $\vol_6$ is unwarped and normalized to $1$. This solution has previously been described in \cite{Silverstein:2004id} (see also \cite{Blaback:2010sj}).
 The smearing has the effect of restoring symmetry in the internal manifold so it reduces to an exact product space
of two spheres. This means that $\e^{2D} = \sin^2\rho$ and all other warp factors and functions are constant. Therefore we find that $F_5$ vanishes 
and the function $\lambda$ introduced in equation \eqref{fluxes} takes a constant value $\lambda = 1$. This corresponds precisely 
to the combined 3-form flux
\be
G_3 = F_3 - i\e^{-\phi} H~,
\ee
being imaginary self-dual, i.e. $\star_6 G_3 = -i G_3$. The size of the 3-form flux is fixed by \eqref{F5bianchi} to be
\be
Q =  g_s |F_3|^2~. 
\ee
The charge conjugated solution would have $Q=-g_s|F_3|^2$ and $\lambda=-1$ which corresponds to anti-ISD fluxes. The expression for the \emph{warped} cosmological constant is
\be
\e^{-2A}\Lambda = -\f14 Q~,
\ee
which shows that if the brane charge $Q$ would decrease, the total vacuum energy would increase.
So already here we notice that a decay process which eliminates $Q$ against $M$ such that the tadpole cancelation 
condition \eqref{tadpole} is still satisfied cannot occur.

\subsection{Compact AdS solutions}
The fully smeared solutions make it clear that compactness is only possible when the worldvolume of the brane is AdS. Flat solutions are necessarily non-compact. It is well-known that for non-compact solutions there is no relation between the CC of the base space and the fluxes.  In compact solutions the size of the CC is determined by the energy, that in turn is determined by the fluxes and the branes. In this subsection we look at compact AdS solutions.

We derive the relation between the CC and the fluxes using the results
of \cite{Gautason:2013zw}. This computation
requires the gauge potential $C_4$. We can choose a gauge for $C_4$ for which the external part, $C_4^\text{ext}$, vanishes at the position  of 
the branes, at $\rho = 0$, 
\be\label{c4potential}
C_4 = B\w C_2 - (\alpha - \alpha_0)\vol_4~.
\ee
Recall that volume forms such as $\vol_4$ are always defined without warp factors. The equation for the cosmological constant of the external spacetime 
can be expressed as follows
\be
\Lambda = \f{1}{4V_6}\left[ N_\text{D3} S_\text{loc} + \f{1}{V_4}\int H\w\left(\e^{-\phi}\star_{10}H + F_3\w C_4^\text{ext}\right)\right]~,
\ee
where $V_4$ and $V_6$ are ``volumes'' of the external and internal spaces, defined as follows
\be
V_4 = \int \vol_4,\qquad V_6 = \int\star_{6}~ \e^{2A}~.
\ee
The gauge choice for $C_4$ is such that, on-shell, the first term in the bracket drops out and we are left only with the flux integral in the second
term. This reduces to 
\bea\label{CC}
\Lambda = \f{\alpha_0}{4V_6} \int H\w F_3~ = -\f{Q\alpha_0}{4V_6}~,
\eea
where we made use of \eqref{tadpole}. Crucially this shows that the function $\alpha$ takes a nonvanishing value at the position of a D3-brane,
this means that $\lambda$ which relates $F_3$ and $H$ in \eqref{fluxes} has the asymptotic behaviour close to the brane:
\be\label{blowup}
\lambda \xrightarrow[\rho\to 0]{} -\f{4V_6\Lambda}{Q}~\e^{-4A}~.
\ee
This obviously blows up since the warp factor $\e^{2A}$ vanishes in the vicinity of the brane. Combined with the fact that $F_3$ is constant 
3-form flux, this leads to the, by now well-known fact that the energy density of $H$ has a singular behaviour close to the brane.

In deriving (\ref{CC}) we have assumed that the solution is compact, although no proof for this exists, apart from the observation that the fully smeared solution is compact. To show that compact solutions can exist one should numerically evaluate the coupled ODE's from appendix \ref{secondordereq}, which we have not done and leave for future investigation.   We verify in Appendix \ref{susy}, that unlike the AdS$_7$ solution from anti-D6 branes, the compact AdS$_4$ solution fom anti-D3 branes cannot be supersymmetric.

\subsection{Non-compact flat solutions}
Compactness enforces the solutions to be AdS but once we give up those two conditions we can consider flat solutions. Flat solutions are interesting from the point of view of antibrane SUSY-breaking \cite{Maldacena:2001pb, Kachru:2002gs, Kachru:2003aw}. Antibrane SUSY-breaking, at least for anti-D3 branes, can be studied explicitly in the Klebanov-Strassler (KS) throat \cite{Klebanov:2000hb}. The KS throat is supersymmetric and regular at the tip and therefore it makes a perfect background to add the singular SUSY-breaking source at the tip. The full solution is out of reach and most likely will remain out of reach, but when the anti-D3 branes are smeared over the tip, the equations of motion, that describe the backreaction become ODE's and approximate solutions have been found \cite{Bena:2009xk,Bena:2011hz}. Close to the tip the details of the singular flux clumping are nicely captured by a much simpler background \cite{Bena:2013hr}, which is the T-dual to the flat anti-D6 solution \cite{Blaback:2011pn}. The only difference from our previous Ansatz are the curvature of the metric factors
\begin{equation}\label{metric2}
 \dd s^2 = \e^{2A}\dd s^2_{\text{Mink}_4} + \e^{2B}\dd s^2_{\mathbb{T}^3/S^3} + \e^{2C}\Bigl(\dd \rho^2 + \e^{2D}\dd \Omega^2\Bigr)\,,
\end{equation}
where $\dd s_{\text{Mink}_4}^2$, $\dd s_{\mathbb{T}^3/S^3}^2$ and $\dd \Omega^2$ are the metrics for Mink$_4$, the three-torus $\mathbb{T}^3$ or the 3-sphere $S^3$ and two-sphere $S^2$ respectively. The solutions with the torus factor can be obtained from T-duality of the anti-D6 solution. When it comes to the physics of flux clumping there is no real difference between the solutions with the torus factor and the solutions with the $S^3$ factor. The solutions with the $S^3$ factor are however more insightful since there is a more explicit picture for brane-flux annihilation in that case \cite{Kachru:2002gs}.

Also here it can easily be demonstrated that $\lambda$ blows up and hence there is singular flux clumping \cite{Bena:2013hr}. When it comes to the local physics associated to the flux clumping, this model captures exactly the same physics as the model with anti-D3's smeared over the tip of the $S^3$ in KS; for instance it can be shown the polarisation potential for spherical D5 branes in both cases is almost identical \cite{Bena:2012vz}. In the next section we compute the polarisation potential both for the polarisation into D5 and NS5 branes.

Since non-compactly we can decouple the size of the CC from the brane charges and the flux quanta, we are free to chose a value for the CC. It is not necessary to take it to be exactly zero. The main result of this paper is that when the CC is set by the brane charges and fluxes, polarisation will occur. When the CC is parametrically different, which can be done for non-compactifications, or in KKLT-type scenarios\footnote{In KKLT type scenarios the CC can de decoupled from the KK scale due to orientifolds and quantum corrections.} \cite{Kachru:2003aw}, then it does not occur and the flux singularity remains unresolved.

\section{Spherical 5-branes}\label{Myers}
Whenever one considers $p$-branes in background with $F_{6-p}$-flux there is the possibility that the $p$-branes polarise into a spherical $(p+2)$-brane carrying $(p+2)$-brane dipole charge and the original $p$-brane monopole charges induced by gauge fluxes on the worldvolume  \cite{Myers:1999ps}. In the case at hand one expects two polarisation channels: a spherical D5 wrapping a contractible $S^2$ inside the $M_3$ that is threaded by $H$  and a spherical NS5 wrapping a contractible $S^2$ inside the $S^3$ filled with $F_3$ (denoted $S^3_F$).  For a compact solution, $M_3$ takes
the form of a three-sphere which we denote by $S^3_H$. S-duality interchanges the roles of D5 and NS5 as well as the fluxes $F_3$ and $H$. So performing an S-duality effectively smears the D3 branes over $S_H^3$ and localises them inside $S^3_F$. We therefore expect the fully localised solution (that is localised on $S^3_H\times S_F^3$) to polarise into a  web of $(p,q)$ 5-branes as in \cite{DHoker:2007xy}. In what follows we discuss both channels.

\subsection{D5 polarisation}
\begin{figure}
\begin{center}
\includegraphics[width=0.5\textwidth]{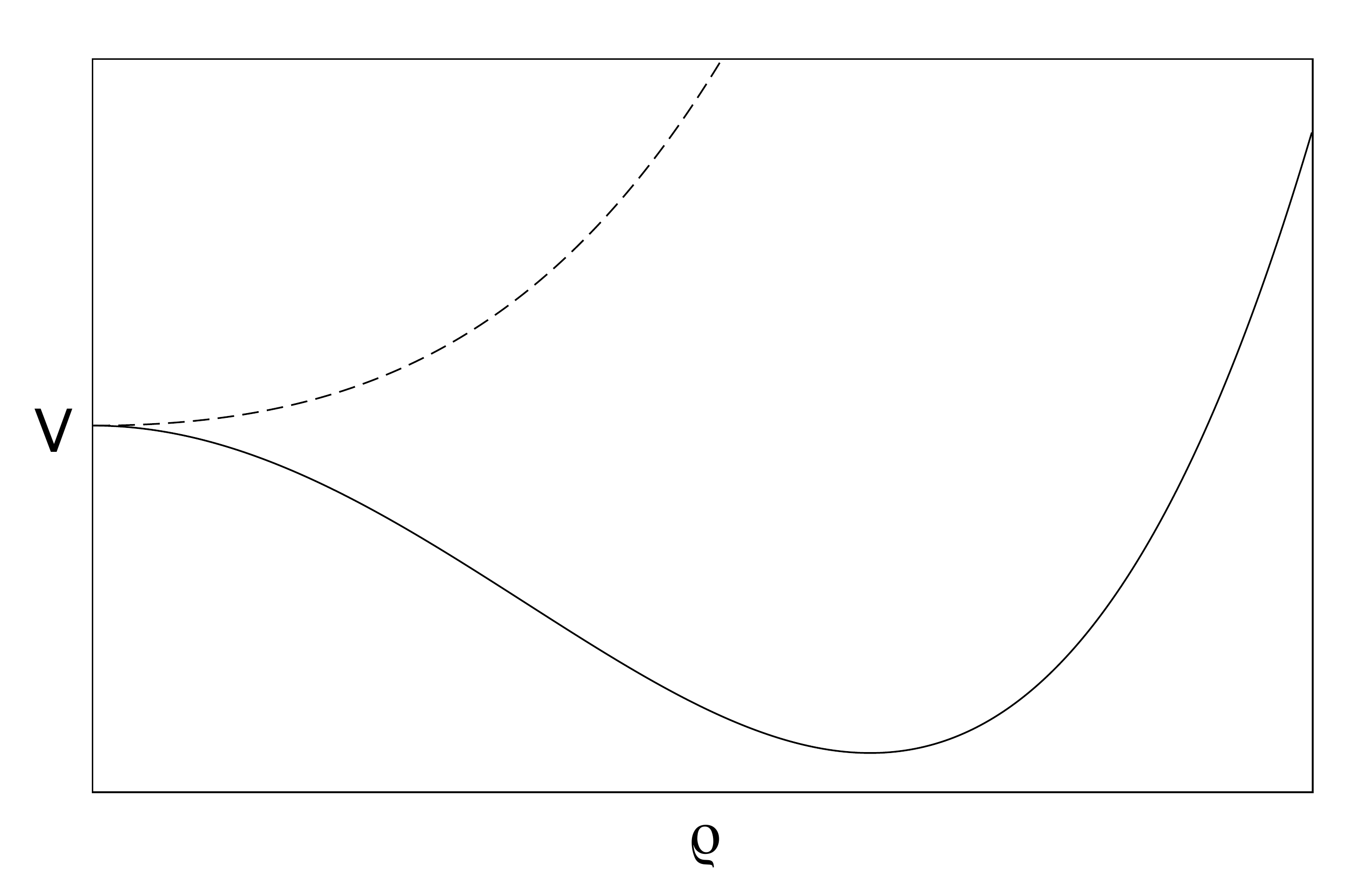}
\caption{\label{D5potential} \small{The potential $V$ (\ref{potentialD5}) for probe D5 branes in the background of D3 branes. The dashed line shows the 
polarisation potential when the external space is flat while the solid line displays the potential for AdS
space, for which the probes are tachyonic near $\bar\rho = 0$.  A stable state exists at a finite distance away from the tip.}}
\end{center}
\end{figure}

There are generically two ways of showing that a D$p$-brane polarizes to a D$(p+2)$-brane; either by considering the non-abelian D$p$-brane action as in
\cite{Myers:1999ps}, or by considering the probe action of a D$(p+2)$-brane in the D$p$-brane background \cite{Polchinski:2000uf}. In this paper we take
the latter approach. The required terms of the D5-brane action in Einstein frame are
\be\label{d5action}
S_\text{D5} = \mu_5 \int \left\{ -\e^{-\phi}\sqrt{-\det(\e^{\phi/2} G - \cal{F})} - C_6 + \cal{F}\w C_4\right\}~, 
\ee
where $\cal{F} = B -  F$. We will take $F = n\pi\vol_2$ and expand the D5-brane action for large $n$ for which polarisation 
is preferred and then look at the  behaviour close to $\rho = 0$. Instead of working through the computation we state the result and in section \ref{genppolarisation} we present a more general computation for $Dp$-branes polarising into $D(p+2)$-branes, which includes this case by putting $p=3$. For AdS$_4$ external space we find
\be\label{potentialD5}
V \propto \left(4\Lambda + \f12k_0^2\right)\bar\rho^2 - 2k_0\bar\rho^3 + 3\bar\rho^4~,
\ee
where $\bar\rho$ is a dimensionless distance from the D3 branes and is defined in section \ref{genppolarisation}. 
The cosmological constant $\Lambda$ is normalized to 
$-3$ for AdS external space and vanishes of course if the external space is flat. The numerical
value $k_0$ is directly related to $\Lambda$ but with a proportionality factor that depends on the details of the solution.
We can however estimate the value of $k_0$ by smearing for which we find $k_0 = \sqrt{6}$ (cf. section \ref{genppolarisation}) 
and the potential takes the form in figure \ref{D5potential}. 

What we find is that the D5 polarisation occurs for the AdS solutions, but that it cannot happen for the non-compact Minkowski solutions. This is in perfect agreement with what has been found for the D6 solutions of \cite{ Blaback:2011nz, Apruzzi:2013yva} as shown in \cite{Bena:2012tx, Junghans:2014wda} (see also \cite{Apruzzi:2013yva, Gaiotto:2014lca}).

The polarisation into (meta-)stable spherical D5 branes smoothens out the singular pile-up of the three-form fluxes. In other words, $\lambda$ remains finite throughout the solution. The only singularities are the expected ones in the metric and form fields that comes from the localised charge and tension of (spherical) branes. This smoothening can be deduced quite easily by repeating the computation of the size of $\lambda$ at the boundary of the spherical D5 brane. One simply needs to use the near-brane expansion of a 5-brane that is smeared over 3-directions. This computation is completely analogues to what has been done in \cite{Bena:2012tx} (equations 2.16 and 2.17).

\subsection{NS5 polarisation and flux decay}

Our AdS$_4$ solution is constructed by smearing the D3 branes over $S^3_F$ which makes it unclear how brane polarisation can proceed inside $S^3_F$. For the polarisation into D5 branes to occur it was necessary to have a solution localised in $M_3$ because the localised solution backreacts in such a way that the profiles of the background supergravity fields induce a minimum in the D5 polarisation potential at a finite size for the spherical 5-brane. But the ``dual'' channel does not need localised branes, one just has to keep in mind that the brane polarisation computation is a \emph{probe} computation and probes are localised. It turns out that the probe computation parallels those done in \cite{Kachru:2002gs, Blaback:2012nf, Danielsson:2014yga}). 

Whereas the physics of the D5 polarisation channel is to resolve the flux clumping, the physics of the NS5 channel is brane-flux decay \cite{Kachru:2002gs}. If the D3 branes polarise into NS5 branes that either tunnel or move perturbatively to the North Pole of the $S^3_F$ one can verify that its monopole charge has shifted to $M-p$ instead of $-p$ D3 charges. The interpretation of this is that $M$ units of 3-form flux materialised into $M$ physical D3 branes that consequently annihilate with the $p$ anti-D3 branes to leave $M-p$ D3 branes in the process.

The NS5 potential is calculated by considering the worldvolume action of a NS5 brane (treated as a probe) in the 
background of D3 branes and fluxes. The main difference to the computation done in \cite{Kachru:2002gs} has to do with the Wess-Zumino (WZ) action for the NS5 brane. The complete NS5 brane action for $C_0=0$ takes the form\footnote{The sign if the WZ action is fixed by demanding that a spherical NS5 brane induces a D3 brance charge,
i.e. that the sign of the $\int F\w C_4$ matches the sign of the WZ action for the D3 branes above.} 
\be
-\mu_{\text{NS5}} \int \e^{-2\phi}\sqrt{-\det\left(\e^{\phi/2}G - \e^{\phi}{\cal F}\right)} + \mu_{\text{NS5}}\int \left(B_6 + {\cal F}\w C_4\right)~,
\ee
The world volume field strength $F$ enters through the combination $\mathcal{F} = F - C_2$.
To evaluate the WZ action we need to find $B_6$. We have
parametrized $H$ as $-\lambda \e^{\phi}\star_6 F_3$ and the dual flux is defined as $H_7 = \e^{-\phi}\star H$. The gauge potential for $H_7$ can be defined through the Bianchi identity $\d H_7 = - F_5\w F_3$.
Evaluating the right hand side for our Ansatz we determine
\be
H_7 = \dd B_6 - C_4\w F_3~.
\ee
From this we find $B_6$ by writing
\be
\dd B_6 = \e^{-\phi}\star_{10} H + C_4\w F_3 = (\lambda\e^{4A} -\alpha+\alpha_0) \vol_4\w F_3 = \alpha_0 \vol_4\w F_3\,,
\ee
such that
\be
B_6 = \alpha_0\vol_4\w C_2~\,,
\ee
which implies that the WZ Lagrangian takes the form
\be
B_6 + {\cal F}\w C_4 = \left(\alpha C_2 - (\alpha-\alpha_0)F\right)\w\vol_4~.
\ee
We also need an expression for $C_2$ that correctly reproduces the constant $F_3$-flux via $F_3 = \dd C_2$
\be
C_2 = \f{M}{2}\left(\psi - \f12 \sin(2\psi)\right)\vol_2~,
\ee
where $\psi$ is used as the third Euler angle on the smeared three-sphere. Remember that the D3 branes are localised 
on $M_3$ at $\rho=0$ (cf. equation \eqref{metric}). The polarisation potential depends on $\psi$ but takes a different form 
depending on the position on 
$M_3$. We will denote the potential by $V_{\rho}(\psi)$.
Finally we let $F_2 = \pi p~\vol_2$ where $p$ sets the D3 brane charge of the probe.
The full polarisation potential is obtained by dividing the NS5 action by $(-\mu_{\text{NS5}} M)$ and relevant volume factors, 
and the result is
\bea\label{NS5bigpotential}
V_{\rho}(\psi) &=& \e^{4A}\sqrt{\f{1}{M^2}\e^{4B-\phi}\sin^4\psi + 
\f{1}{4}\left(2\pi\f{p}{M}-\psi+\f12\sin(2\psi)\right)^2}\nonumber\\
 &&- \f{\alpha}{2}\left(\psi - \f12 \sin(2\psi)\right) - \pi(\alpha-\alpha_0)\f{p}{M}.
\eea
This potential is valid for either the AdS or flat D3 brane solutions discussed above. The main difference when analysing the 
potential lies in the fact that the D3 branes in AdS polarize immediately to D5 branes.

\begin{figure}
\begin{center}
\includegraphics[width=0.5\textwidth]{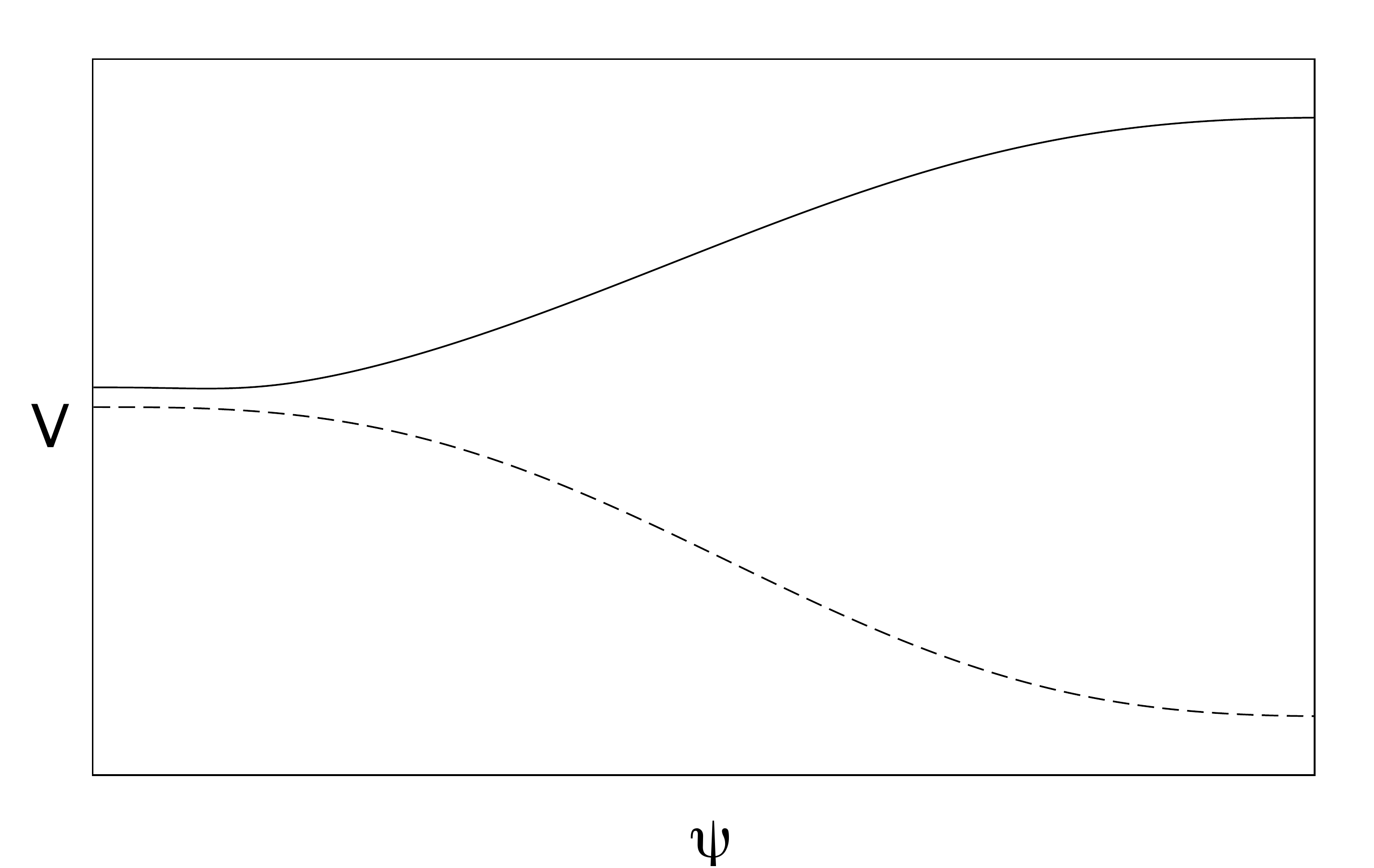}
\caption{\label{NS5fig}The polarisation potential for spherical NS5 branes. The dashed line is a plot of the potential 
\eqref{NS5bigpotential} for flat external space and at $\rho=0$. The solid line is evaluated for AdS external space with
$\lambda_\text{max} = 1/2$.}
\end{center}
\end{figure}

Let us discuss the potential in the flat case for which we have shown that the D3 branes do \emph{not} polarise into D5 branes. Even though $\e^{4A}$ vanishes as $\rho\to 0$, $\alpha$ stays constant.
This means that the first term in \eqref{NS5bigpotential} vanishes as $\rho\to 0$ but the second term does not.
The potential therefore reduces to
\be
V_{\rho = 0}(\psi) =  -\f{\alpha_0}{2}\left(\psi - \f12 \sin(2\psi)\right)~.
\ee
The number $\alpha_0$ is finite and positive, just as for the compact AdS solution. This can be understood from studying 
the $F_5$ Bianchi identity which leads to a strong constraint on the form of $\alpha$ when combined with D3 brane 
boundary conditions and the asymptotic behaviour of the fields far away from the D3s. The analysis is completely 
analogous to the one done in \cite{Blaback:2011pn} in the case of D6 branes. Hence the NS5 potential for flat branes indicate a perturbative brane-flux decay as shown in picture \ref{NS5fig}.

For an AdS$_4$ external space the situation is different, the polarisation of the D3 branes to D5's regularizes the
singularity in $\lambda$ such that the both terms in equation \eqref{NS5bigpotential} play a role. Brane-flux decay depends on the relative size of the two terms in the potential. The first term effectively pulls the 
NS5 probe towards the D5 branes while the second term pushes the probe away. We have seen that for the fully smeared
solution brane-flux decay cannot occur for a very simple reason; the energy of the vacuum would increase. We expect the
same result to hold in the partially localised case (and in the fully localised case). In order to confirm this we would
have to scan the potential $V_\rho$ for all values of $\rho$ and show that the gravitational pull of the first term outweighs
the electromagnetic push of the second one. It would then be enough to find an upper bound on the function $\lambda$ and
show that it is less than the smeared value $\lambda = 1$. For the AdS vacua, we have plotted the expected qualitative behavior for the NS5 potential in figure \ref{NS5fig}.

\section{Anti-D$p$ solutions}\label{generalDpsolutions}
The D3 brane solutions discussed so far belongs to a class of D$p$-brane solutions studied first in \cite{Blaback:2010sj}, where
the smeared limit and flat limit were explored. Here we study the setup in more generality. 

The metric takes the form
\be\label{generalmetric}
\dd s^2 = \e^{2A}\dd s^2_{\text{AdS}_{p+1}} + \e^{2B}\dd s^2_{S^{6-p}} + \e^{2C}\dd s^2_{M_3},\quad\text{and}\quad \dd s^2_{M_3} = \dd \rho^2 + \e^{2D}\dd \Omega^2~,
\ee
where $\dd s_{\text{AdS}_{p+1}}^2$, $\dd s_{S^{6-p}}^2$ and $\dd \Omega^2$ are the metrics for AdS$_{p+1}$, the $(6-p)$-sphere $S^{6-p}$ and 
the two-sphere $S^2$ respectively. Once again we choose the metrics such that the corresponding unwarped Ricci tensors take the canonical form
\be
\hat{R}_{\mu\nu} = \Lambda~ \hat{g}_{\mu\nu}~,\quad \hat{R}_{mn} = (5-p)~ \hat{g}_{mn}~,\quad \hat{R}_{ij} = \hat{g}_{ij}~.
\ee
The hatted variables denote the unwarped quantities, and
\be
\dd s^2_{\text{AdS}_{p+1}} = \hat{g}_{\mu\nu} \dd x^\mu\dd x^\nu~, \qquad \dd s^2_{S^{6-p}} = \hat{g}_{mn} \dd x^m\dd x^n~, \quad
\dd \Omega^2 = \hat{g}_{ij} \dd x^i\dd x^j~.
\ee
The warp factors $A$, $B$, $C$ and $D$ are only functions of the coordinate $\rho$ and so the Einstein equations reduce to ordinary differential 
equations for the warp factors. 
We parametrize our fluxes as follows:
\bea
H &=& -\lambda \e^{\f{p+1}{4}\phi} \star_{9-p} F_{6-p}~,\nonumber\\
F_{6-p} &=& M{\vol}_{S^{6-p}}~,\label{generalfluxes}\\ 
F_{8-p} &=& \e^{-(p+1)A - \f{p-3}{2}\phi}\star_{9-p}\dd\alpha~.\nonumber
\eea
The functions $\lambda$ and $\alpha$ as well as the dilaton are assumed, like the warp factors, to depend only on the $\rho$ coordinate.
The volume form ${\vol}_{6-p}$ is the unwarped volume form on the $(6-p)$-sphere.
From the equation of motion for $H$ we determine
\be\label{Heom}
\alpha = -(-1)^p \lambda \e^{(p+1)A + \f{p-3}{4}\phi} = -(-1)^p \lambda\beta~,
\ee
where the function $\beta$ is defined by the second equality.
The only unsatisfied form field equation is the $F_{8-p}$ Bianchi identity, which takes the form
\be\label{generalbianchi}
\dd F_{8-p} - H\w F_{6-p} = N_{\text{D}p}\mu_p \delta_{9-p}~.
\ee

\subsection{The AdS curvature}
Once again we can relate the zero point value of the function $\alpha$ to the AdS curvature via the relation given in \cite{Gautason:2013zw}. For the 
Ansatz specified above we find
\be
\f{2}{p-1}\Lambda = \f{1}{4V_{p+1}}\left[N_{\text{D}p} S_\text{loc} + \f{1}{V_{9-p}}\int H\w\left(\e^{-\phi}\star_{10} H - \sigma(F_{6-p})\w C_{p+1}\right)\right]~,
\ee
from which we can get rid of the first term by a gauge choice for $C_{p+1}$:
\be
C_{p+1} = - (\alpha - \alpha_0)\sigma(\vol_{p+1})~,
\ee
where $\alpha_0$ is the value of $\alpha$ at $\rho=0$. Putting this together we get
\be\label{genCC}
\f{2}{p-1}\Lambda = \f{\alpha_0}{4V_{9-p}}\int H\w F_{6-p} = -\f{ \alpha_0 Q}{4V_{9-p}}~,
\ee
where we used equation \eqref{generalbianchi} to evaluate the integral and
\be
V_{9-p} = \int \star_{9-p}~\e^{(p-1)A}~.
\ee
Since the warped volume $V_{9-p}$ is not known a priori, the equation \eqref{genCC} does not fix the value of $\alpha_0$. 
However, we now know that for non-vanishing cosmological constant, the value of $\alpha_0$ must be a strictly positive
number. This fact enables us to conclude that a singularity is developed in the energy density of $H$ just as for the
special case $p=3$ discussed above. Since the argument is completely analogous we do not repeat it here.

Concerning supersymmetry we explain in Appendix \ref{susy} that compact AdS$_p$ solutions cannot be supersymmetric for $p=3,4,5$, when an anti-D$p$ singularity is assumed at $\rho=0$.

\subsection{Flat solutions}
As with the anti-D3 branes, we can also investigate flat solutions, or solutions with a non-zero cc, but whose value is not fixed by compactness and hence decoupled from the scale set by the brane charges and the fluxes. The  metric Ansatz is simply the generalisation of the Ansatz used for anti-D$3$ branes:
\be\label{generalmetric2}
\dd s^2 = \e^{2A}\dd s^2_{\text{Mink}_{p+1}} + \e^{2B}\dd s^2_{\mathbb{T}^{6-p}/S^{6-p}} + \e^{2C}\dd s^2_{M_3},~.
\ee
These solutions describe anti-Dp branes smeared over the $\mathbb{T}^{6-p}/S^{6-p}$. The solutions with the torus factor can be obtained from T-duality of the anti-D6 solution \cite{Blaback:2011pn}.

\subsection{Brane polarisation}\label{genppolarisation}
We compute the potential for a probe D$(p+2)$-brane in the background of 
$N_{\text{D}p}$ D$p$-branes. By now this is a standard computation that we repeat for completeness and we find agreement with
the results of \cite{Bena:2012tx,Junghans:2014wda} for $p=6$ and \cite{Bena:2012vz} for $p=3$.

The probe action in this case is
\be\label{dp+2action}
S_{\text{D}(p+2)} = -\mu_{p+2} \int \left\{ \e^{-\phi}\sqrt{-\det(\e^{\phi/2} G - \cal{F})} - (-1)^p\sigma(C_{p+3} - \cal{F}\w C_{p+1})\right\}~,
\ee
where as before $\cal{F} = B-F$ and $F$ is the world volume field strength. We take $F = \pi n~\vol_2$ and expand the 
action for large $n$ to obtain
\be\label{generalpotential}
V \propto (\pi n - b)L_{\text{D}p} + (-1)^p\gamma + \f{\beta\e^{\phi+4C+4D}}{2(\pi n - b)}~,
\ee
where 
\be
B = b(\rho) ~\vol_2,\quad C_{p+3} = \gamma(\rho)~ \sigma(\vol_{p+1})\w\vol_2~,
\ee
and
\be
L_{\text{D}p}(\rho) = \beta - (-1)^p(\alpha - \alpha_0) = \beta(1+\lambda) + (-1)^p\alpha_0~.
\ee
The functions $b(\rho)$ and $\gamma(\rho)$ are determined by employing the definition of $H$ and $F_{p+4}$ in terms of their potentials
and comparing to the anzats. The result is 
\bea
b'(\rho) &=& M\alpha\e^{\phi-(p+1)A - (6-p)B + 3C + 2D}~,\nonumber\\
\gamma'(\rho) &=&  \left(\beta^2 - \alpha(\alpha-\alpha_0)\right)M\e^{\phi-(p+1)A - (6-p)B + 3C + 2D}~.
\eea
We will use these equations to determine the behaviour of $b$ and $\gamma$ close to the D$p$-branes. Before we expand the fields
we present a differential equation for $L_{\text{D}p}$ which is obtained by combining the external 
Einstein equation with the dilaton equation and the Bianchi identity \eqref{generalbianchi}. The equation is remarkably simple, and in particular
no source terms appear:
\be\label{mastereq}
\nabla^2L_{\text{D}p} - \beta^{-1}(\nabla L_{\text{D}p})^2 = \beta\left[ (p+1)\Lambda\e^{-2A} + \f{(1+\lambda)^2}{2}\e^{\f{p-1}{2}\phi}|F_{6-p}|^2\right]~.
\ee

We are now in position to expand the fields close to the D$p$-branes so as to obtain an expression for the potential $V$ close to the branes.
We use the standard boundary conditions of the fields close to a D$p$ brane,
\bea
\e^{2A} &\approx& \rho^{\f{7-p}{8}} (a_0 + a_1\rho)~,\nonumber\\
\e^{2B} &\approx& \rho^{\f{-1-p}{8}} (b_0 + b_1\rho)~,\nonumber\\
\e^{2C} &\approx& \rho^{\f{-1-p}{8}} (c_0 + c_1\rho)~,\\
\e^{2D} &\approx& \rho^{2} (1 + d_1\rho)~,\nonumber\\
\e^{2\phi} &\approx& \rho^{\f{p-3}{2}} (f_0 + f_1\rho)~,\nonumber\\
L_{\text{D}p} &\approx& \rho(l_0 + \rho l_1)~,\nonumber
\eea
the last expansion in this list is determined by noting that we chose the gauge for $C_{p+1}$ such that the constant part of $L_{\text{D}p}$ 
vanishes. The constants $a_0,b_0,c_0$ and $f_0$ can be rewritten in terms of the number of Dp branes $N_{\text{D}p}$ and string coupling $g_s$ by studying the flat $p$-brane solutions. The near brane
behaviour of $\alpha$ is given by
\be
\alpha \approx \alpha_0 + \rho\alpha_1~.
\ee
Below we need the first term in the expansion of $\beta$
\be
\beta \approx \rho\beta_0 = \rho f_0^{\f{p-3}{8}}a_0^{\f{p+1}{2}}~.
\ee
Expanding the equation \eqref{mastereq} to leading order we immediately find $l_0=0$ and this implies
\be
\beta_0 = (-1)^p\alpha_1~.
\ee
The next order coefficient is
\be
6l_1 = \f{c_0\beta_0}{a_0}\left((p+1)\Lambda + \f12 (\lambda_0\beta_0 M)^2 f_0^{\f12}a_0^{-p}b_0^{p-6}\right)~.
\ee
Expanding the gauge potentials gives
\bea
b(\rho) &\approx& \f12\rho^2 \alpha_0Mf_0^{\f12}a_0^{-\f{p+1}{2}}b_0^{-\f{6-p}{2}}c_0^{\f32}~,\nonumber\\
\gamma(\rho) &\approx& -\f13\rho^3\alpha_0\alpha_1Mf_0^{\f12}a_0^{-\f{p+1}{2}}b_0^{-\f{6-p}{2}}c_0^{\f32}~.
\eea
Using these results it is straightforward to write down the first few terms of the expansion of the brane potential,
\be\label{genfullpotential}
V \approx \f{\beta_0(\pi n)^3}{6a_0^2 f_0^{\f12}}\left\{\left((p+1)\Lambda + \f12 k_0^2\right)\bar\rho^2
- 2k_0 \bar\rho^3
+ 3\bar\rho^4 \right\}~,
\ee
where 
\be\label{knotdefinition}
\bar\rho = \f{\sqrt{c_0a_0}f_0^{\f14}}{\pi n}\rho\quad \text{and}\quad k_0 = (\alpha_0 M) a_0^{-\f{p}{2}} b_0^{-\f{6-p}{2}}f_0^{\f14}~.
\ee
We now see that the polarisation potential \eqref{genfullpotential} is fully determined as soon as the constant $k_0$ is given.

Besides the extremum at the origin the potential has at most one more extremum at
\be\label{extremum}
\bar{\rho} = \f{k_0}{2} \pm \sqrt{-\f{k_0^2}{12}-\f{(p+1)\Lambda}{3}}~,
\ee
which shows that polarisation occurs in AdS space when
\be\label{polarisationcondition}
-\Lambda\ge \f{k_0^2}{4(p+1)}~.
\ee
One of our key results is that polarisation for solutions where the CC is decoupled from the fluxes and brane charges is 
impossible. This in particular includes the flat solutions which obviously do not polarise by equation \eqref{extremum}.

Let us now estimate $k_0$ for the AdS solutions. The second-order equations of motions can unfortunately not help 
without solving the  system completely. We can however estimate $k_0$ by solving the equation \eqref{mastereq} 
in the fully smeared limit. This amounts to 
putting the derivatives to zero and replacing $\lambda$ with its smeared value \cite{Blaback:2010sj}
\be
\lambda \to \f{p-1}{2}~.
\ee
We then find
\be
\f12 k_0^2 \to \f{p(p-1)^2}{p+1}~,
\ee
which can be inserted into the polarisation potential \eqref{genfullpotential}. 
The potential is easily verified from equation \eqref{polarisationcondition} to allow for a polarisation at a 
finite value of $\bar\rho$.
 
In this section we used two expansions, one for large $n$ and another for small radius $\rho$. Clearly we can choose $\rho$ small  to justify the near brane expansion and $N_{\text{D}p}$ large  to stay within the probe approximation, $n\ll N_{\text{D}p}$. Within this regime a minimum of the D$(p+2)$ potential can be deduced in the following way. Since the WZ term becomes less important when we move away from the anti-D$p$, the DBI term dominates forcing the potential upward. By checking whether the D$(p+2)$ potential decreases away from the point $\rho=0$ we deduce the existence of a minimum. Obviously, a more careful analysis is needed to quantitatively trust the calculation from the previous section up to the minimum of the potential but this is unnecessary to verify brane polarisation.

\section{Conclusion}\label{conclusion}

We have shown that there is a consistent picture for the flux singularities associated with anti-D$p$ solutions that are smeared over $6-p$ compact directions. These solutions come in two types: compact AdS$_{p+1}$ solutions and non-compact flat solutions and both feature singular fluxes that partially screen the antibrane charges. We have found a story similar to what happened for localised anti-D6 branes \cite{Apruzzi:2013yva, Junghans:2014wda, Bena:2012tx}: the flat solutions do not polarise into spherical branes whereas the AdS solutions do. As a consequence the AdS solutions have regular flux clouds in the supergravity limit.  Compact AdS  vacua have a  cosmological constant related to the energy in such a way that brane-flux decay would \emph{increase} the energy. This is opposite to the flat solution, where brane-flux decay lowers the energy. So either flux clouds that are singular at the SUGRA level initiate perturbative brane-flux decay \cite{Blaback:2012nf, Danielsson:2014yga} or brane polarisation has to occur in order to resolve the flux singularity. We have found exactly that.   Concerning the flat solutions we also find consistency with \cite{Blaback:2012nf, Danielsson:2014yga}: the flux clumping is too large and causes perturbative brane-flux decay such that the smooth solution is expected to be time dependent.  This is in agreement with the absence of regular finite temperature solutions \cite{Blaback:2014tfa, Bena:2012ek, Bena:2013hr, Hartnett:2015oda}.  
\begin{figure}
\begin{center}
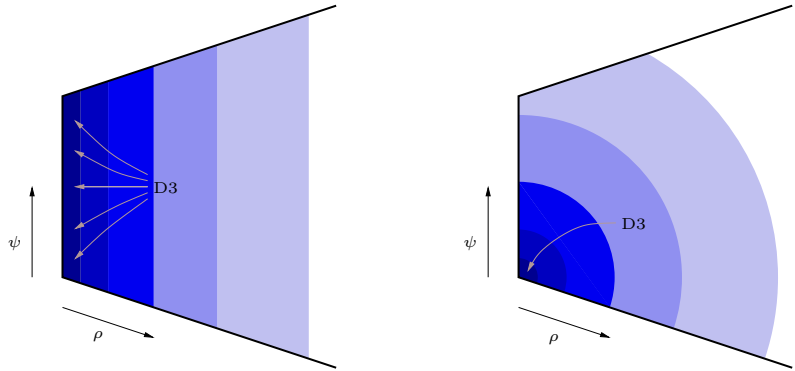
\caption{\label{smearedvslocalized} {\small The difference between the flux clumping between antibranes smeared over the tip of the KS throat (left) and antibranes localised at the South Pole of the tip (right). At every point in the $\psi$-direction we suppressed the two-sphere (with zero size at South and North Pole, $\psi=0,\pi,$ and maximum size in between). The flux clumping on the right is less severe in the middle of the $\psi$ direction, such that the force that pushes the NS5 towards the North Pole is less and could be small enough to create a classical barrier. }}
\end{center}
\end{figure}

An important restriction of our work is the smearing of the antibranes over the internal $S^{6-p}$ or $\mathbb{T}^{6-p}$. It is important to investigate brane polarisation and brane-flux decay for fully localised branes. This becomes especially important in case one looks at backgrounds with very few, or even a single, SUSY-breaking antibrane. For a single antibrane it is clearly not physical to smear the charges over a compact submanifold and it is exactly in the regime of a single brane that it has been argued that the meta-stable states are most likely to exist \cite{Kachru:2002gs, Michel:2014lva}. An important argument here in favor of meta-stability is that the flux clumping effects of \cite{Blaback:2012nf, Danielsson:2014yga} are only relevant at the South Pole of the $S^3$ and once the NS5 brane moves away from the South Pole the forces that push it over the equator are diluted\footnote{We like to thank U. Danielsson, D. Junghans and the authors of \cite{Michel:2014lva} for discussions on that matter.} as depicted in figure \ref{smearedvslocalized}.

At first sight this allows a classical barrier against brane-flux decay and would seem consistent with a recent claim that regular finite temperature solutions can exist if the antibranes are fully localised \cite{Hartnett:2015oda}. If this is the case, it is a good indication that there is no tachyonic mode that describes the onset of brane-flux decay and it would therefore be most important to find a full proof of the claims in \cite{Hartnett:2015oda}.

The sensitivity of antibrane uplifting to instabilities arises through the  use of warped throat that redshifts string-scale energies down to much smaller energy scales. This locally creates a lack of scale-separation and various modes can mix with Kaluza--Klein (KK) modes that become light \cite{Aharony:2005ez}. The flux-clumping instabilities, if present, are an example of this effect since flux gradients correspond to KK modes and mix with the modes that correspond to the NS5 position.  Checking full stability remains therefore a subtle issue for scenarios that are based on antibrane uplifting. One could for instance worry about the tachyonic modes found in \cite{Bena:2014jaa} or even Gregory-Laflamme-like instabilities in the screened anti-D3 brane \cite{Ulfpaper}.

Finally, there is a black hole analogue to antibrane SUSY breaking in flux backgrounds, which are near-extremal micro-state geometries build from meta-stable supertubes \cite{Bena:2011fc, Bena:2012zi}. One could worry whether these constructions also feature the problem of enhanced decay due to flux-clumping like effects. We have good reasons to believe this is not the case and hope to report on this in the near future. 

\subsection*{Acknowledgements}
We would like to thank Iosif Bena, Ulf Danielsson, Alessandro Tomasiello, Daniel Thompson, Joseph Polchinski, Andrea Puhm and especially Nikolay Bobev for useful discussions.
TVR is supported by a Pegasus fellowship and by the Odysseus programme of the FWO.  BT is aspirant FWO. We also acknowledge support from the European Science Foundation Holograv Network.

\appendix
\section{Notation and conventions}\label{Notation}

The bulk type II action takes the form
\be\label{bulkaction}
S = \int \star_{10}\left\{R - \f12 |\dd\phi|^2 - \f12 \e^{-\phi}|H|^2 - 
\f14 \sum_{n} \e^{\f{5-n}{2}\phi}|F_n|^2\right\}~,
\ee
where $R$ is the curvature scalar of the Einstein frame metric $G$ with mostly plus signature. The kinetic terms for the dilaton $\phi$, the NSNS 3-form 
$H$ and the RR forms $F_n$ are written using the short hand notation
\be
|\omega_p|^2~\star_{10}1 = \star_{10}\omega_p\w\omega_p = \f{1}{p!}G^{M_1N_1}\cdots G^{M_pN_p} \omega_{M_1\cdots M_p}\omega_{N_1\cdots N_p}~\star_{10}1~,
\ee
where
\be
\omega_p = \f{1}{p!}\omega_{M_1\cdots M_p} \dd X^{M_1}\w\cdots \w\dd X^{M_p}~,
\ee
is any $p$-form. Notice that we are using the democratic formulation of \cite{Bergshoeff:2001pv} which means that all RR field strengths $F_n$ with 
$n=1,3,5,7$ and $9$ in type IIB and $n=0,2,4,6,8$ and $10$ in type IIA appear in the action. In this way the
bulk action does not contain any Chern-Simons terms but a duality relation between the RR fields,
\be\label{dualitycondition}
\e^{\f{5-n}{2}\phi}F_n = \star_{10}\sigma(F_{10-n})~,
\ee
must be imposed by hand on-shell. The reversal operator $\sigma$ has been introduced to simplify many equations in the following, it 
does only introduce a sign depending on the degree of the form it acts on, i.e.
\be
\sigma(\omega_p) = (-1)^{\f{p(p-1)}{2}}\omega_p~.
\ee
Including the localized action for a anti-D$p$ brane
\be\label{locaction}
S_\text{loc} = -\mu_p \int_{N_{p+1}}  \left\{ \e^{\f{p-3}{4}\phi}\star_{p+1}1 + (-1)^p\sigma(C_{p+1})\right\}~,
\ee
we see that the Bianchi identity for $F_{8-p}$, the Einstein equation and the dilaton equation acquire correction due to the presence of the branes. In 
the localized action we have already made use of the fact that the worldvolume field strength $\cal{F}$ vanishes in the setup we consider. The fields appearing in the D-brane action are understood as the pull-backs of their bulk counterparts. For reference we present the modified
Bianchi identity for $F_{8-p}$
\be\label{F8-pbianchi}
\dd F_{8-p} - H\w F_{6-p} = \mu_p \delta_{9-p}(N_{p+1})~,
\ee
where $\delta_{9-p}(N_{p+1})$ denotes the $p+1$-form with delta distribution support on the brane worldvolume, i.e. 
\be
\delta_{9-p}(N_{p+1}) = \star_{9-p}1~ \delta(N_{p+1})~.
\ee

\section{Second-order equations}\label{secondordereq}
In this appendix we present the second-order differential equations for the Ansatz in section \ref{antiD3sol}.

The $F_5$ Bianchi identity implies
\begin{equation}
(\alpha'e^{3B+C+2D-4A})'=\lambda M^2 e^{\phi-3B+3C+2D} - N_{\text{D}3}\mu_3\delta_6~.
\end{equation}
We use a prime to denote a derivative with respect to $\rho$. The dilaton equation gives
\begin{equation}
\phi'' + (4A +3B+C+2D)'\phi' = \frac{1}{2}M^2e^{\phi+2C-6B}(1-\lambda^2)~.
\end{equation}
We present the Einstein equation in the trace reversed form 
\be
R_{MN} = \hat{T}_{MN}~.
\ee
The Ricci tensor is
\begin{eqnarray}
R_{\mu\nu} &=&  -e^{-2 C} \left(4 e^{2 A} A'^2+e^{2 A} \left(3 B'+C'+2 D'\right) A'+3 e^{2 C}+e^{2 A} A''\right)e^{-2A}g_{\mu\nu}\,, \nonumber\\
R_{ij}&=& e^{-2 C} \left(-3 e^{2 B} B'^2-4 e^{2 B} A' B'-e^{2 B} \left(C'+2 D'\right) B'+2 e^{2 C}-e^{2 B} B''\right)e^{-2B}g_{ij}\,, \nonumber\\
R_{\rho\rho}&=&-4 A'^2+4 C' A'-3 B'^2-2 D'^2+3 B' C'-2 C'D'\nonumber\\
&&-4 A''-3 B''-2 C''-2 D''\,,\nonumber\\
R_{ab}&=& \Bigl(1-e^{2 D} \left(C'+D'\right)^2-4 e^{2 D} A' \left(C'+D'\right)-3 e^{2D} B' \left(C'+D'\right)\nonumber\\ 
&&-e^{2D} \left(D'^2+C'D'+C''+D''\right)\Bigr)e^{-2C-2D}g_{ab} \,.
\end{eqnarray}
The components of the trace-reversed energy-momentum tensor are
\begin{eqnarray}
\hat{T}_{\mu\nu} &=&\tfrac{-1}{8}g_{\mu\nu}\Bigl((1+\lambda^2)M^2 e^{\phi-6B} + 2e^{-8A-2C}(\alpha'^2) + 2\mu_3N\delta\Bigr)~,\nonumber\\
\hat{T}_{ij} &=&\tfrac{1}{8}g_{ij}\Bigl(M^2(3-\lambda^2)e^{\phi-6B} + 2(\alpha')^2e^{-8A-2C} \Bigr)~,\nonumber\\
\hat{T}_{\rho \rho} &=& \tfrac{1}{2}\phi'^2 +\tfrac{1}{8}e^{2C+\phi - 6B}(3\lambda^2 - 1)M^2 -\tfrac{1}{4}(\alpha)'^2e^{-8A}
~,\nonumber\\
\hat{T}_{ab} &=&\tfrac{1}{8}g_{ab}\Bigl(M^2(3\lambda^2-1)e^{\phi-6B} + 2(\alpha')^2e^{-8A-2C} \Bigr)~.
\end{eqnarray}

\section{Supersymmetric AdS}\label{susy}
The BPS-equations for our AdS solutions are (\ref{generalmetric}-\ref{generalfluxes}): 
\bea
16(\nabla\phi)^2&=& - \gamma^{-2}M^2\left(4\beta^{-2}\alpha^2 - (p-1)^2\right)\nonumber\\
&&+ (p-3)^2\beta^{-2}(\nabla \alpha)^2~,\nonumber\\
(\nabla (4A+\phi))^2&=&-16e^{-2A}  +  \left[\gamma^{-2}M^2+ \beta^{-2}(\nabla \alpha)^2\right] \nonumber\\
\left(4\e^{-B} \pm \nabla (4B+\phi)\right)^2 &=& \left[\gamma^{-2}M^2+ \beta^{-2}(\nabla \alpha)^2\right]\nonumber\\
(\nabla(3\phi-4C-4D))^2 &=&\left( 4e^{-D-C} - (p-2)\beta^{-1}\nabla \alpha\right)^2\nonumber\\
&& +  (p-2)^2\gamma^{-2}M^2~,
\eea
where
\bea
\beta &=& \e^{(p+1)A + \f{p-3}{4}\phi}~,\nonumber\\
\gamma &=& \e^{(6-p)B - \f{p-1}{4}\phi}~.
\eea
In addition to these equations we must supplement also the Bianchi identity for $F_{8-p}$ in equation \eqref{generalbianchi}.
We have verified that this system reproduces the one presented in \cite{Apruzzi:2013yva} for $p=6$ for which no $B$ warp
factor is present. For $p=5$  the system is also modified as the third equation takes the form
\be
(\nabla (4B+\phi))^2 = \left[\gamma^{-2}M^2+ \beta^{-2}(\nabla \alpha)^2\right]~.
\ee
This is due to the fact that for general $p$ the metric has a factor $S^{6-p}$ which for $p=5$ is simply a circle and
the associated Ricci tensor must vanish. 

The first two BPS equations can be combined so that $\nabla\alpha$ does not appear,
\be
(p-3)^2(\nabla (4A+\phi))^2 - 16(\nabla \phi)^2 = -16(p-3)^2\e^{-2A} + 4 \gamma^{-2}M^2\left(\lambda^2-p+2\right)~.
\ee
Expanding this equation around a D$p$ singularity and comparing with analogous expansion of the
following equation of motion:
\be
-4p(p-3)\e^{-2A} - 4(p-3)\nabla^2A+(7-p)\nabla^2\phi = - 2\gamma^{-2}(\beta^{-2}\alpha^2-1)~,
\ee
we obtain an expression for the constant $k_0$ which was introduced in section \ref{genppolarisation}
\be
\f12 k_0^2 = \f{(p-4)(p-3)^2}{(p-5)}~.
\ee
This equation has important consequences because for $p=3$ and $p=4$ we see that $\alpha_0$ 
vanishes (cf. equation \eqref{knotdefinition}). The fact that $\alpha_0$ vanishes is however in direct
contradiction with our previous result that $\alpha_0$ is proportional to the non-zero CC. We must conclude
that the assumption we made when deriving the relation between $\alpha_0$ and CC; that the internal space
is compact, is not true for $p=3,4$. This result only holds when we assume a D$p$ singularity at $\rho=0$, 
a compact solution with no brane at the pole might of course exist.

\bibliographystyle{utphys}
{\footnotesize
\bibliography{refs}} 
\end{document}

%% file: smevsloc.pdf_t
\begin{picture}(0,0)%
\includegraphics{smevsloc.pdf}%
\end{picture}%
\setlength{\unitlength}{1657sp}%
\begingroup\makeatletter\ifx\SetFigFont\undefined%
\gdef\SetFigFont#1#2#3#4#5{%
  \reset@font\fontsize{#1}{#2pt}%
  \fontfamily{#3}\fontseries{#4}\fontshape{#5}%
  \selectfont}%
\fi\endgroup%
\begin{picture}(11658,5466)(526,-5494)
\put(1801,-5011){\makebox(0,0)[lb]{\smash{{\SetFigFont{6}{7.2}{\rmdefault}{\mddefault}{\updefault}{\color[rgb]{0,0,0}$\rho$}%
}}}}
\put(8551,-5056){\makebox(0,0)[lb]{\smash{{\SetFigFont{6}{7.2}{\rmdefault}{\mddefault}{\updefault}{\color[rgb]{0,0,0}$\rho$}%
}}}}
\put(9631,-3391){\makebox(0,0)[lb]{\smash{{\SetFigFont{6}{7.2}{\rmdefault}{\mddefault}{\updefault}{\color[rgb]{0,0,0}D3}%
}}}}
\put(541,-3661){\makebox(0,0)[lb]{\smash{{\SetFigFont{6}{7.2}{\rmdefault}{\mddefault}{\updefault}{\color[rgb]{0,0,0}$\psi$}%
}}}}
\put(7291,-3661){\makebox(0,0)[lb]{\smash{{\SetFigFont{6}{7.2}{\rmdefault}{\mddefault}{\updefault}{\color[rgb]{0,0,0}$\psi$}%
}}}}
\put(2701,-2851){\makebox(0,0)[lb]{\smash{{\SetFigFont{6}{7.2}{\rmdefault}{\mddefault}{\updefault}{\color[rgb]{0,0,0}D3}%
}}}}
\end{picture}%